\documentclass[twocolumn,twocolappendix]{aastex7}

\usepackage{amsmath}
\usepackage{amssymb}
\usepackage{CJK}
\usepackage{booktabs}
\usepackage{graphicx}
\usepackage{caption}
\usepackage{subcaption}
\usepackage{graphicx}

\newcommand{\RNum}[1]{\uppercase\expandafter{\romannumeral #1\relax}}

\begin{document}

\title{The Balmer Absorber Strength Problem of Little Red Dots}

\begin{CJK*}{UTF8}{gbsn}

\author[0000-0003-2488-4667]{Hanpu Liu (刘翰溥)}
\affiliation{Department of Astrophysical Sciences, Princeton University, 4 Ivy Lane, Princeton, NJ 08544, USA}
\email[show]{hanpu.liu@princeton.edu}

\author[0000-0001-9185-5044]{Eliot Quataert}
\affiliation{Department of Astrophysical Sciences, Princeton University, 4 Ivy Lane, Princeton, NJ 08544, USA}
\email{quataert@princeton.edu}

\author[0000-0001-9269-5046]{Bingjie Wang (王冰洁)}
\thanks{NHFP Hubble Fellow}
\affiliation{Department of Astrophysical Sciences, Princeton University, 4 Ivy Lane, Princeton, NJ 08544, USA}
\email{bjwang@princeton.edu}

\author[0000-0002-5612-3427]{Jenny E. Greene}
\affiliation{Department of Astrophysical Sciences, Princeton University, 4 Ivy Lane, Princeton, NJ 08544, USA}
\email{jgreene@astro.princeton.edu}

\author[0000-0002-2624-3399]{Yan-Fei Jiang (姜燕飞)}
\affiliation{Center for Computational Astrophysics, Flatiron Institute, New York, NY 10010, USA}
\email{yjiang@flatironinstitute.org}

\begin{abstract}
Little Red Dots (LRDs) commonly show Balmer absorption features. Some observed LRD spectra have weaker absorption in H$\alpha$ than in H$\beta$, in apparent contradiction to atomic physics according to standard attenuation models. This tension is naturally resolved when one considers emission from the absorbing medium itself, which is ignored in the standard model:   when the Balmer absorption is optically thick, a non-negligible source function could fill in the H$\alpha$ absorption trough and reduce its observed absorption equivalent width (EW). With proof-of-concept photoionization models we show that the source function of H$\alpha$ could be systematically stronger than that of H$\beta$. This is due to radiative trapping preferentially populating the $n=3$ level of neutral hydrogen, which operates efficiently when the medium is optically thick to H$\alpha$ but not very optically thick to Paschen-$\alpha$. Such line-dependent source functions could lead to significant corrections to the expected EW and selectively weaken H$\alpha$ absorption, consistent with observations. Quantitatively, this depends on the distance of the absorbing medium and its internal structure, providing a new diagnostic of the environment around LRDs. Future modeling of the absorbers may also potentially open a new avenue for estimating the mass of the central engine. 

\end{abstract}


\section{Introduction} 
\label{sec:intro}

Little Red Dots (LRDs) are a distinctive group of extragalactic objects discovered with the James Webb Space Telescope. Their compact morphology and broad, intense Balmer emission lines resemble quasars. However, various aspects differ sigfinicantly from known Type-I active galactic nuclei (AGNs). Typical LRDs show weak or absent X-ray emission \citep{Yue2024,Ananna2024,Maiolino2025,Sacchi2025}, little short-term variability \citep{Kokubo2025,ZhangZ2025,Burke2025,Stone2026,LiuZ2026}, weak to undetectable high-ionization lines such as He\,\textsc{ii} or C\,\textsc{iv} \citep{Lambrides2026,Tang2025,Wang2026,Zucchi2026}, and a red rest-optical color without strong hot or cold dust emission \citep{Williams2024,Akins2024,Setton2025,Xiao2025,Casey2025}. Their broad Balmer emission lines show very strong decrements \citep{Killi2024,Brooks2025,Nikopoulos2026,Yanagisawa2026a} and commonly have absorption features \citep{Matthee2024,Matthee2026,Lin2026,Juodzbalis2026,Yanagisawa2026b}. A subgroup of LRDs present a Balmer break that is difficult to explain with a stellar population \citep{Labbe2024,Ma2025,Ji2025,Naidu2025,deGraaff2025}. These observational pecularities suggest physical conditions different from typical AGNs.  

To account for some of these phenomena, many current models invoke a high column density of gas associated with the central engine (e.g., a gas cocoon, \citealt{Inayoshi2025,Naidu2025,Rusakov2026,Sneppen2026}; a thermalized atmosphere, \citealt{Kido2025,Begelman2025,Nandal2026,LiuH2026,Gentile2026}; or geometrically thick accretion or outflows, \citealt{Lambrides2026,Liu2025,Naidu2026,Madau2026}). In particular, a hydrogen volume density $n_{\rm H}>10^8{\rm~cm^{-3}}$ with a high column density and near-unity covering factor is inferred to explain the Balmer absorption features (Balmer absorbers hereafter) without overproducing He\,{\textsc{i}} absorption in a case study \citep{Juodzbalis2024}. The same dense gas could explain the Balmer decrement by gas radiative transfer effects \citep{Yan2026} and is potentially also responsible for producing the Balmer break \citep{Inayoshi2025}. In other words, observations suggest unique gas conditions in LRDs, which introduce an array of physical effects that are absent or less manifest in the broad line region of typical AGNs. 

The Balmer absorbers of LRDs are of particular interest to this work. Notably, the absorption features can show increasing depth for higher order transitions: in a few individual LRD cases, H$\alpha$ is weaker than H$\beta$ in terms of the equivalent width (EW) or the best-fit optical depth \citep[e.g.,][]{DEugenio2026a,DEugenio2025,DEugenio2026b,Torralba2026,Matthee2026,Chen2026,Lin2026,Wang2026b}. This is the opposite of what one would expect from their oscillator strengths.

Main-sequence A-type stars show precisely such a trend, with absorption strengthening all the way to H$\delta$ or H$\epsilon$ \citep[also noted in, e.g.,][]{DEugenio2026a}. Importantly, the lines are optically thick: the flux in the absorption trough is no longer a simple attenuation of the incident radiation but is instead dominated by the locally generated radiation, or, more precisely, the source function. In a classical photoionized medium, the source function is negligible compared to the emission line or the continuum, which justifies modeling absorbers as pure attenuation. However, it remains unclear whether the same approach applies to the inferred unique gas conditions of LRDs. We will show that the pure-absorption assumption may break down, and that considering the source function naturally explains the weak absorption of H$\alpha$. 

This paper outlines a simple analysis of the optically thick Balmer absorbers of LRDs. Among the LRDs that show weak H$\alpha$ absorption relative to H$\beta$, we take one example as an observational point of comparison, RUBIES-EGS-49140 \citep[RUBIES-49140 hereafter,][]{Wang2024,Lambrides2025,DEugenio2025,Wang2026b}. In Section~\ref{sec:problem}, we identify the problem of weak H$\alpha$ absorption in a simple analytical framework and heuristically point out the source function as a solution to this problem. In Section~\ref{sec:microphysics}, we elaborate on the microphysical conditions for producing the weak absorption by photoionization modeling. Then, in Section~\ref{sec:curve_of_growth_source}, we apply the modeling results to reconcile the observed absorber strengths with theoretically expected source functions. We summarize our results and discuss the physical and observational implications in Section~\ref{sec:summary}. Throughout this paper, we use the term ``absorber'' to denote the observed absorption feature rather than the physical medium that causes the absorption, and we always quote wavelengths and luminosities in the rest frame unless otherwise noted. We adopt the \citet{PlanckColl2020} cosmological parameters, with $\Omega_m =0.310, \Omega_\Lambda=0.690$ and $H_0=67.66{\rm~km~s^{-1}~Mpc^{-1}}$.

\section{The problem: H$\alpha$ absorption is surprisingly weak}
\label{sec:problem}
Several works have noted that, at least for a fraction of LRDs, the measured H$\alpha$ absorption is surprisingly weak compared to H$\beta$ \citep[e.g.,][]{DEugenio2026a,Torralba2026,Matthee2026,Chen2026,Wang2026b}. We quantify this comparison in the form of the analytical ``curve of growth'', i.e., the dependence of the equivalent width of an absorption feature as a function of the line-center optical depth (e.g., \citealt{Draine2011}; see, e.g., \citealt{Lin2026}, for application to LRDs). Throughout this work, we assume that the absorbing medium is located outside the continuum photosphere and the broad-line region but within the narrow-line region, so the base of the absorption feature is the sum of the continuum and the broad line. For notation, we consistently use unprimed indices (e.g., $n$ or $n\ell$) for the upper atomic level and primed indices (e.g., $n'$ or $n'\ell'$) for the lower atomic level. 

For bound-bound transitions of the hydrogen atom, the optical depth at the line center is given by
(\citealt{Draine2011}; see also \citealt{Inayoshi2025})
\begin{align}
    \tau_{n',n} &= {\sqrt{\pi}e^2\over m_ec}{\lambda_{n',n}f_{n',n}N_{n'}\over b} \nonumber\\
    &= 8.69~{\lambda_{n',n}\over\lambda_{2,4}}{f_{n',n}\over f_{2,4}}{N_{n'}\over10^{15}{\rm cm^{-2}}}\left({b\over100{\rm~km\,s^{-1}}}\right)^{-1}\,, \label{eq:optical_depth_center}
\end{align}
where $n$ and $n'$ are the principal quantum numbers of the upper and lower levels (so $\tau_{2,3}$ and $\tau_{2,4}$ correspond to H$\alpha$ and H$\beta$), $e$ and $m_e$ are the electron charge and mass, $c$ is the speed of light, $\lambda_{n',n}$ and $f_{n',n}$ are the wavelength and oscillator strength of a given absorption transition $n'\to n$, $N_{n'}$ is the column density of the $n'$-th level, and $b$ is proportional to the one-dimensional velocity dispersion, $b=\sqrt{2}\sigma_v$. For a given line, $\tau_{n',n}\propto N_{n'}$. Stimulated emission is ignored in Equation~(\ref{eq:optical_depth_center}), which is only used for analytical estimates in this section. Our numerical calculations later shows a $<10\%$ correction due to stimulated emission in all cases that we consider. For simplicity, Equation~(\ref{eq:optical_depth_center}) also omits the index for the azimuthal quantum number $\ell$ by assuming that subshells at $n=2$ are statistically populated due to collisional $\ell$-mixing. We relax this assumption and develop an $\ell$-resolved analysis in Section~\ref{sec:microphysics}; there, in our photoionization models, the $n=2$ sublevels are verified to be $\ell$-mixed. 

\begin{figure*}
    \centering
    \includegraphics[width=0.99\linewidth]{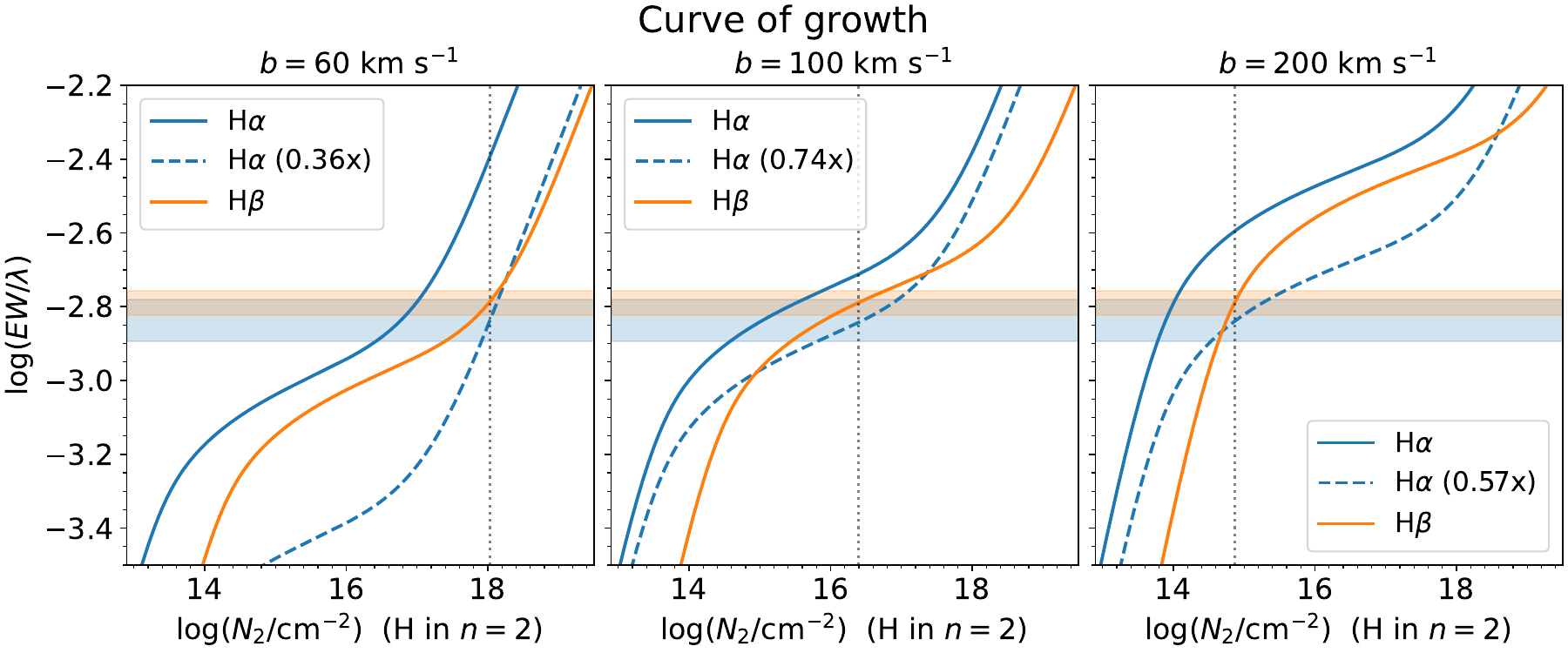}
    \caption{The EWs of H$\alpha$ and H$\beta$ absorption as a function of $n=2$ column density (known as the curve of growth). The three panels correspond to different microturbulence velocities. Shaded horizontal bands are the observationally estimated EWs for H$\alpha$ (blue) and H$\beta$ (orange) for RUBIES-49140 with $\pm1\sigma$ uncertainties. The standard curve of growth predicts $(EW/\lambda)_{\rm H\alpha}>(EW/\lambda)_{\rm H\beta}$ for all $N_2$, contrary to observation; downward shifts of the curve of growth (conceptually illustrated with dashed curves) due to a nonzero source function could reconcile the tension.  }
    \label{fig:curve_of_growth}
\end{figure*}

Some example curves of growth for H$\alpha$ and H$\beta$ are visualized in Figure~\ref{fig:curve_of_growth}. For a given transition and a given $b$, the curve has three regimes in order of increasing optical depth: the optically thin regime, the saturated regime (the Doppler core is optically thick, but the damping wing remains optically thin), and the damping regime (the wing due to natural damping becomes optically thick). In all three regimes, the EW increases with $N_2$, although the slope is shallower in the intermediate regime. Therefore, an observed EW uniquely maps to a value of $N_2$. 

This poses a tension with observations. The curve of growth of H$\beta$ is always lower than that of H$\alpha$, mainly due to the difference in oscillator strength: $f_{2,3}=0.6411$ and $f_{2,4}=0.1194$ \citep{Wiese2009}. Therefore, for a given absorbing medium, i.e., for a given $N_2$ and $b$, one always expects $(EW/\lambda)_{{\rm H}\alpha}>(EW/\lambda)_{{\rm H}\beta}$. This is opposite to observation. For example, for RUBIES-49140, $\log(N_2/{\rm cm^{-2}})=16.4$ at $b=100{\rm~km~s^{-1}}$ explains the observed $\log(EW/\lambda)_{{\rm H}\beta}=-2.79\pm0.03$ \citep[measurements from][]{Wang2026b}, indicated by the gray dotted line in the middle panel. However, for the same conditions to explain H$\alpha$, the curve of growth of H$\alpha$ has to be scaled by a factor of 0.74. 
A partial covering factor as in the standard attenuation model \citep[e.g.,][]{Juodzbalis2024} does not alleviate the tension because it will scale the H$\alpha$ and H$\beta$ curves of growth simultaneously without changing their ratios; we discuss this further in Section~\ref{sec:summary}. We note that our curves of growth  ignore collisional broadening relevant in stellar atmospheres, notably Stark broadening and hydrogen self-broadening \citep[e.g.,][]{Fuhrmann1993,Barklem2000}. They would become significant only at densities higher than typically considered for LRDs: the former at an electron density $n_e>10^{12}{\rm~cm^{-3}}$, and the latter at a neutral hydrogen density $n_{\rm HI}>10^{15}{\rm~cm^{-3}}$. Even if any of these effects dominate, they would somewhat allieviate but not resolve the tension. More details can be found in Appendix~\ref{app:stark}. 

The downward scaling of 0.74 for H$\alpha$ to explain the inverted H$\alpha$-to-H$\beta$ EW ratio in Figure~\ref{fig:curve_of_growth} seems arbitrary. However, the standard curve of growth assumes that the source function of the optically thick absorption line is negligible. We now show that accounting for a finite line source function changes this result. Specifically, the formal solution of radiative transfer implies
\begin{equation}
    \boxed{I(\lambda) = CI_{\rm in}(\lambda)e^{-\tau(\lambda)}  + (1-C)I_{\rm in}(\lambda) + CS(\lambda)(1-e^{-\tau(\lambda)})}\,, \label{eq:attenuation_model}
\end{equation}
where $I(\lambda)$ is the spectral flux density after the absorption at a given wavelength $\lambda$, $I_{\rm in}(\lambda)$ is that before the absorption (including the continuum and the broad line), $0\leq C\leq1$ is the covering factor, and $S(\lambda)$ is the source function, which is in general different for different transitions. Throughout this work, we use the source function defined per unit wavelength $(\rm erg\,cm^{-2}\,s^{-1}\,\AA^{-1}\,sr^{-1})$. The standard attenuation model excludes the last term in Equation~(\ref{eq:attenuation_model}). However, when non-negligible, it is the key term that can resolve the tension between H$\alpha$ and H$\beta$ absorption strengths. The source function plays a vital role in areas such as stellar and supernova spectral analysis \citep[e.g.,][]{Jeffery1990}.

With Equation~(\ref{eq:attenuation_model}), the EW of the absorber is
\begin{align}
    EW &\equiv \int\left[1-{I(\lambda)\over I_{\rm in}(\lambda)}\right]d\lambda\nonumber\\
    &= \int C\left[1-{S(\lambda)\over I_{\rm in}(\lambda)}\right](1-e^{-\tau(\lambda)})d\lambda\nonumber\\
    &= C\left[1-\left\langle{S\over I_{\rm in}}\right\rangle\right]EW^*\,, \label{eq:EW}
\end{align}
where 
\begin{equation}
    \left\langle{S\over I_{\rm in}}\right\rangle\equiv {\int {S(\lambda)\over I_{\rm in}(\lambda)}(1-e^{-\tau(\lambda)})d\lambda \over \int(1-e^{-\tau(\lambda)})d\lambda} \label{eq:mean_source_over_intensity}
\end{equation}
is the source function divided by the incoming intensity averaged over the absorber profile, and
\begin{equation}
    EW^* \equiv \int(1-e^{-\tau(\lambda)})d\lambda \label{eq:EW_star}
\end{equation}
is the EW from the standard curve of growth, which assumes $S(\lambda)=0$ and $C=1$. We will later use the subscript $\langle S/I_{\rm in}\rangle_{2,3}$ and $\langle S/I_{\rm in}\rangle_{2,4}$ to denote the quantity for H$\alpha$ and H$\beta$ respectively. For simplicity, the attenuation model here assumes a uniform source function throughout the absorbing medium. In the more general case, the averaging in Equation~(\ref{eq:mean_source_over_intensity}) should also account for depth variations of the source function, which we will return to in Section~\ref{subsec:cloudy}. 

Equation~(\ref{eq:EW}) indicates that, when in general $S(\lambda)>0$, the EW is simply scaled by a factor $C[1-\langle S/I_{\rm in}\rangle]$. This offers a natural physical interpretation of the dashed curves in Figure~\ref{fig:curve_of_growth}: we simply require $\langle S/I_{\rm in}\rangle_{2,3} >\langle S/I_{\rm in}\rangle_{2,4}$ to shift down the H$\alpha$ curve of growth relative to H$\beta$. As a numerical example, assuming a fiducial $b=100{\rm~km\,s^{-1}}$ and $C=1$, if $\langle S/I_{\rm in}\rangle_{2,3}=0.26$ and $\langle S/I_{\rm in}\rangle_{2,4}\lesssim0.1$ (so the H$\beta$ curve is barely shifted), then the absorbing medium will explain both EWs at a consistent $\log(N_2/{\rm cm^{-2}})=16.4$ (gray dotted line in the middle panel in Figure~\ref{fig:curve_of_growth}). Alternatively, assuming $b=200{\rm~km\,s^{-1}}$ and $C=1$, $\langle S/I_{\rm in}\rangle_{2,3}=0.43$ and $\langle S/I_{\rm in}\rangle_{2,4}\lesssim0.1$ will allow an absorbing medium at $\log(N_2/{\rm cm^{-2}})=14.9$ to explain both EWs. We may also allow H$\beta$ to shift: for example, when $b=100{\rm~km\,s^{-1}}$ and $C=1$, if $\langle S/I_{\rm in}\rangle_{2,3}=0.47$ and $\langle S/I_{\rm in}\rangle_{2,4}=0.20$ (not visualized), then the consistent solution will move to $\log(N_2/{\rm cm^{-2}})=17.4$. The column density here is broadly in agreement with previous estimates for other LRDs with absorbers \citep[e.g.,][where $\log(N_2/{\rm cm^{-2}})\sim15$]{Juodzbalis2024,Juodzbalis2026}. 

\section{Microphysics of the absorbing medium}
\label{sec:microphysics}
The main message from Section~\ref{sec:problem} is that weaker absorption in H$\alpha$ than in H$\beta$ can be naturally explained by a nonzero source function, with $\langle S/I_{\rm in}\rangle_{2,3} >\langle S/I_{\rm in}\rangle_{2,4}$. We now explore what microphysical conditions may produce this. From now on, we will use double indices $n\ell$ for individual atomic subshells, whereas single indices still denote $\ell$-level-unresolved quantities as before. 

\subsection{Source function from level populations}
A Balmer line involves several $\ell$-resolved transitions, although their transition energies are close enough to be unresolvable in LRDs. For example, for H$\alpha$, all permitted transitions $n'\ell'\to n\ell$ where $n'=2, n=3$ contribute to the spectral line. Therefore, we define the observationally meaningful line source function to be $\ell$-unresolved but account for these individual contributions:
\begin{equation}
    S_{n',n} = {2hc^2\over\lambda_{n',n}^5}{\sum_{\ell\ell'} A_{n\ell,n'\ell'}n_{n\ell}\over\sum_{\ell\ell'} A_{n\ell,n'\ell'}({g_{n\ell}\over g_{n'\ell'}}n_{n'\ell'}-n_{n\ell})}\,, \label{eq:source_function}
\end{equation}
where $h$ is the Planck constant, $A_{n\ell,n'\ell'}$ is the Einstein A coefficient for the transition $n\ell\to n'\ell'$, $g_{n\ell}$ is the statistical weight, and $n_{n\ell}$ is the number density of hydrogen at the level $n\ell$. The sums in the numerator and denominator are respectively proportional to the total emissivity and the total stimulation-corrected opacity from individual transitions. We have assumed complete frequency redistribution for the absorption and emission profiles, which is a good approximation at least in the Doppler core of the line \citep{Jefferies1960}. According to Equation~(\ref{eq:source_function}), the source function is directly determined by the upper-level and lower-level population densities: it increases with the former and decreases with the latter, whereas all other terms are constant for a given transition. 

How does this compare to the pre-absorption intensity $I_{\rm in}$? The latter is related to the observable rest-frame pre-absorption luminosity:
\begin{equation}
    I_{\rm in} = {L_{\rm in}\over4\pi^2 R^2}\,, \label{eq:radius}
\end{equation}
where $R$ is the distance from the absorbing medium to the central engine. Fitting to RUBIES-49140 in \citet{Wang2026b} suggests a peak luminosity (broad line plus continuum) $L_{\rm in,2,3}=1.64^{+0.06}_{-0.05}\times10^{42}{\rm~erg~s^{-1}~\AA^{-1}}$ for H$\alpha$ and $L_{\rm in,2,4}=(2.63\pm0.012)\times10^{41}{\rm~erg~s^{-1}~\AA^{-1}}$ for H$\beta$. We use these peak luminosities as a representative value for the luminosity across the absorber, which is justified by the small observed absorber offset and FWHM compared to the FWHM of the broad line. With Equation~(\ref{eq:radius}), we can estimate
\begin{equation}
    \left\langle{S\over I_{\rm in}}\right\rangle_{2,n}\approx 4\pi^2 R^2{\bar{S}_{2,n}\over L_{{\rm in},2,n}}\,,\quad n=3,4\,, \label{eq:radius_inference}
\end{equation}
where $\bar{S}_{2,n}$ is the source function averaged over the line of sight through the medium, weighted according to the optical depth; we will define this formally later in Section~\ref{subsec:cloudy}. Here, we have assumed that the continuum does not significantly contribute to the source function across the absorber profile. This aligns with our overall assumption that the continuum and the broad line have already formed before entering the absorbing medium. 

The distance $R$ is highly uncertain. However, our goal for now is to compare $\langle S/I_{\rm in}\rangle_{2,3}$ and $\langle S/I_{\rm in}\rangle_{2,4}$. Dividing them cancels out $R$:
\begin{equation}
    {\langle S/I_{\rm in}\rangle_{2,4}\over\langle S/I_{\rm in}\rangle_{2,3}} \approx {L_{\rm in,2,3}\over L_{\rm in,2,4}}{\bar{S}_{2,4}\over \bar{S}_{2,3}}\,. \label{eq:source_function_ratio}
\end{equation}
In our fiducial numerical example in Section~\ref{sec:problem}, we gave $\langle S/I_{\rm in}\rangle_{2,3}=0.26$ and $\langle S/I_{\rm in}\rangle_{2,4}\lesssim0.1$. In light of Equation~(\ref{eq:source_function_ratio}), this will require $\bar{S}_{2,4}/\bar{S}_{2,3}\lesssim0.06$. 

\subsection{\textsc{cloudy} modeling}
\label{subsec:cloudy}

We now use photoionization models to show that a low $\bar{S}_{2,4}/\bar{S}_{2,3}$ is achievable under plausible conditions of the LRD absorbing medium. We caution that the models and calculations in this subsection are oversimplified; they are intended to be a proof of principle rather than a constraining model for observations. 

We compute physical conditions of the Balmer absorbing medium with the code \textsc{cloudy} \citep[version C25, last described by][but with modifications; see below]{Gunasekera2025}. Each model is a one-dimensional gas layer of constant hydrogen density $n_{\rm H}=10^{10}{\rm~cm^{-3}}$, with a closed geometry (i.e., the ``sphere'' command in the code\footnote{This command enforces that photons cannot escape via reflection off the illuminated surface of the gas layer; despite the name of the command, light propagation is still assumed to be plane-parallel.}, justified by the high covering factor of the absorbing medium). We also explored $n_{\rm H}=10^8{\rm~cm^{-3}}$ and $10^{12}{\rm~cm^{-3}}$ and found similar results. We use the solar abundance set of \citet{Asplund2009} with all metals scaled to $0.1Z_\odot$. We apply a fixed microturbulence of $b=100{\rm~km~s^{-1}}$. The thickness of the gas layer is controlled by the total hydrogen column density $N_{\rm H}$. 

The radiation illuminating the absorbing medium remains poorly understood. For simplicity, we assume a blackbody at an effective temperature $T_{\rm eff}=20000{\rm~K}$ for the incident radiation. The effective temperature is lower than conventionally used for AGNs as the weak He\,\textsc{ii} emission of LRDs suggests a lack of high-energy incident photons \citep{Wang2026}. The intensity is parameterized by the surface flux of hydrogen-ionizing photons, $\log(\Phi_{\rm H}/{\rm cm^{-2}\,s^{-1}})$. This is related to the more commonly used ionization parameter $U$ by $\Phi_{\rm H}=Un_{\rm H}c$. We choose $\Phi_{\rm H}$ in favor of $U$ because the former is at zeroth order proportional to $N_2$ independent of $n_{\rm H}$ \citep{Ferland1979}. We explore a range of $16\leq\log(\Phi_{\rm H}/{\rm cm^{-2}\,s^{-1}})\leq18$ and $20\leq\log(N_{\rm H}/{\rm cm^{-2}})\leq27.2$.

We modify the release version of \textsc{cloudy} to improve the treatment of resonance-line pumping by the incident continuum. The code by default adopts a self-shielding formula originally developed for molecular destruction in the photodissociation region \citep{Federman1979}, which assumes that photons at the far wing of a line are destroyed when captured. This is inapplicable to resonance lines, e.g., Ly$\alpha$, where photons at far wings are predominantly coherently scattered; the latter effect is considered in the escape probability calculation of \textsc{cloudy} but not in its default shielding formula. In other words, the default setup assigns different probabilities for a photon to penetrate from the gas surface to a given depth and for a photon emitted at that depth to reach the surface. This violates the reversibility of photon paths for resonance lines. Therefore, we instead evaluate the shielding of each line with the same escape-probability function that governs its emission. \textsc{Cloudy} has already implemented this with an optional command ``set continuum shielding Pesc'', which nevertheless still left the treatment of Ly$\alpha$ asymmetrical. Following the format of that existing implementation, we modify the code to match the shielding function of Ly$\alpha$ to its fiducially adopted escape probability formula \citep{Hummer1980} as well and apply it to all calculations in the main body of this work. We validate this modification and discuss its implications for our production runs and for wider applications of \textsc{cloudy} in Appendix~\ref{app:cloudy_modification}, and we have made our modification and tests publically available and reproducible in a Zenodo repository \citep{Liu2026software}. 

To diagnose each \textsc{cloudy} calculation, we extract the level populations of atomic hydrogen and evaluate the source functions for the $2\to3$ and $2\to4$ transitions (Equation~\ref{eq:source_function}) at each location. Then, we calculate a depth-averaged source function defined by
\begin{equation}
    \bar{S}_{n',n}={{\int S_{n',n}}e^{-\tilde\tau_{n',n}}d\tilde\tau_{n',n}\over\int e^{-\tilde\tau_{n',n}}d\tilde\tau_{n',n}}\,, \label{eq:mean_source_function}
\end{equation}
where $\tilde\tau_{n',n}$ is the outside-in stimulation-corrected optical depth, and the integral is performed across the entire gas layer. The weighting by $e^{-\tilde\tau_{n',n}}$ accounts for contributions from the source function at different optical depths to the emergent radiation. We note that we do not use the optical depth at the line center $\tau_{n',n}$ in Equation~(\ref{eq:optical_depth_center}). The observational line fitting is strongly driven by the wings of the absorber, and thus the inferred EW is mainly determined by the source function far from the line center, where the optical depth is much lower. We choose to measure the optical depth at $3b$ away from the line center, with $\tilde\tau_{n',n}=\exp({-3^2})\tau_{n',n}$. The value of $3b$ corresponds to an FWHM of $6\sqrt{\ln2}\,b=500{\rm~km~s^{-1}}$, comparable to the observationally measured value for absorbers \citep{Wang2026b}; using $2b$ or $4b$ instead would not change our qualitative conclusion. This evaluation can be done independently for each wavelength under complete frequency redistribution, where the line source function is independent of wavelength (Equation~\ref{eq:source_function}). 

\begin{figure}
    \centering
    \includegraphics[width=0.99\linewidth]{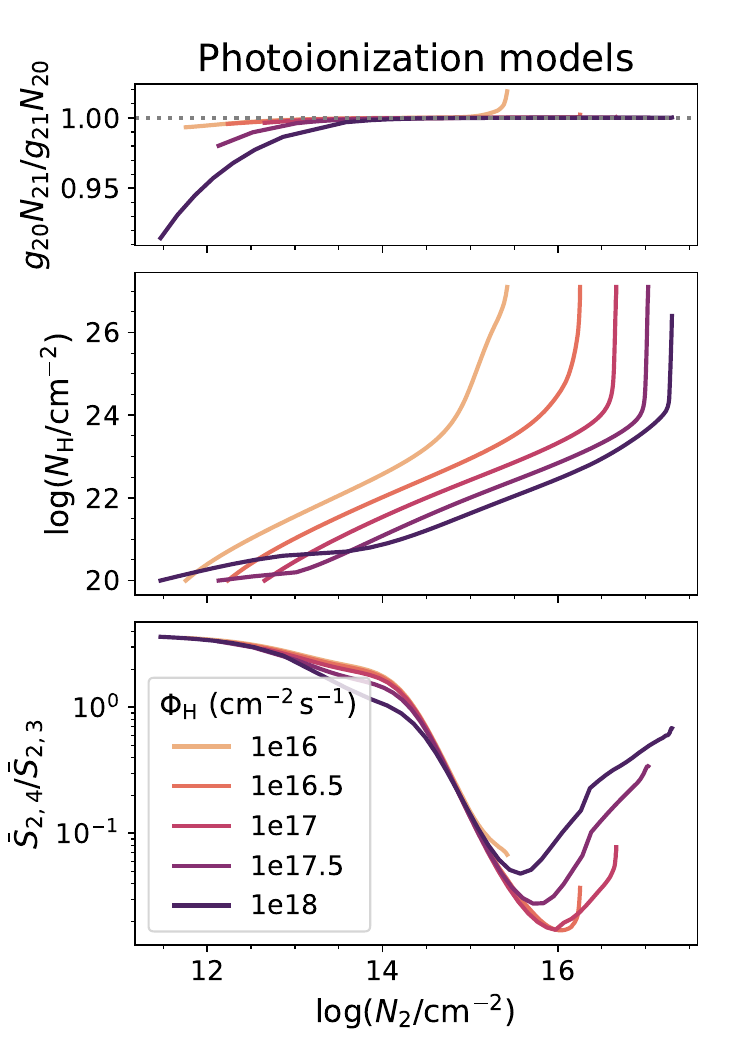}
    \caption{Photoionization calculations. Each point on each curve corresponds to a \textsc{cloudy} absorbing medium model at a given $(N_{\rm H}, \Phi_{\rm H})$. The horizontal axis is the measured total column density of the $n=2$ population $N_2$. Top: ratio between the column densities of 2p and 2s; $g_{2s}N_{2p}/g_{2p}N_{2s}=1$ indicates that 2s and 2p are $\ell$-mixed. Middle: relation between the input paramters $N_{\rm H}$ and the measured $N_2$. They monotonically correlate with each other and follows $N_2\propto N_{\rm H}$ at low column density. Bottom: ratio of the mean source functions of H$\beta$ to H$\alpha$, $\bar{S}_{2,4}/\bar{S}_{2,3}$. Models at intermediate $N_2$ reaches a valley of $\bar{S}_{2,4}/\bar{S}_{2,3}\sim10^{-2}$, which facilitates weak H$\alpha$ absorption relative to H$\beta$.}
    \label{fig:cloudy_grid}
\end{figure}

Figure~\ref{fig:cloudy_grid} shows the results of \textsc{cloudy} calculations. Different from convention, we present the measured quantity $N_2$ at the horizontal axes instead of the input parameter $N_{\rm H}$, which is shown on the vertical axes in the middle panel. This arrangement makes the plots more similar to the curve of growth and improves physical interpretability. 

We first briefly comment on the status of $\ell$-mixing in our models. We mentioned in Section~\ref{sec:problem} that our analytical curve of growth assumed statistically populated subshells at $n=2$. The top panel in Figure~\ref{fig:cloudy_grid} diagnoses this by comparing the column densities of the 2p and 2s sublevels, $N_{2p}$ and $N_{2s}$: if they are $\ell$-mixed, then $g_{2s}N_{2p}/g_{2p}N_{2s}=1$. This turns out to be accurate within 10\% for all cases here. This validates the self-consistency between the analytical and numerical models at least for the parameter space that we have explored. On the other hand, atomic levels at $n\geq3$ are not always $\ell$-mixed (not shown in the figure), so we still deem it necessary to evaluate the source function by indivual sublevels (Equation~\ref{eq:source_function}). 

Next, we focus on the middle panel. All curves show a positive trend between $N_2$ and $N_{\rm H}$: at low $N_{\rm H}$, the two correlate approximately linearly, but $N_2$ saturates at high $N_{\rm H}$. Across different curves, higher $\Phi_{\rm H}$ mostly leads to higher $N_2$ for the same $N_{\rm H}$. We directly inspect the converged transition rates among the atomic levels and find that creation of the $n=2$ population is mainly due to Lyman pumping by the incident continuum and the recombination cascade, while destruction is dominated by Ly$\alpha$ escape at low $N_{\rm H}$ and by a combination of two-photon continuum decay ($2s\to1s$) and collisional de-excitation at high $N_{\rm H}$. The main creation channels of the $n=2$ population are radiative in nature, so their depth-integrated rates are approximately proportional to $\Phi_{\rm H}N_{\rm H}^0$. On the destruction side, the depth-integrated Ly$\alpha$ escape rate is inversely proportional to its escape probability, which aymptotically scales as $N_1^{-1}\simeq N_{\rm H}^{-1}$; the two-photon decay and collisional de-excitation do not strongly depend on the optical depth. Equilibrium then suggests that $N_2\propto \Phi_{\rm H}N_{\rm H}$ at low $N_{\rm H}$ and that $N_2\propto \Phi_{\rm H}$ at high $N_{\rm H}$, which is broadly consistent with the trends in the figure. Overall, $N_{\rm H}$ serves as a control parameter for the target variable $N_2$, the latter being directly relevant to the curve of growth for H$\alpha$ and H$\beta$. 

Most pertinent to the EW problem is the behavior in the ratio $\bar{S}_{2,4}/\bar{S}_{2,3}$ (bottom panel). For each $n_{\rm H}$, the low-$N_2$ and the high-$N_2$ ends show a large ratio of $\bar{S}_{2,4}/\bar{S}_{2,3}\sim10^0$. However, the ratio plunges at intermediate column densities of $N_2\sim10^{15{\text{--}17}}{\rm~cm^{-2}}$, in many cases reaching $\bar{S}_{2,4}/\bar{S}_{2,3}\sim10^{-2}$. This trough concurs with the range of $N_2$ expected from the curve of growth to explain the observed EWs of RUBIES-49140 (Figure~\ref{fig:curve_of_growth}). We devote the next subsection to studying the mechanism behind this trend. 

\subsection{The mechanism of weak H$\alpha$ absorption}
\label{subsec:mechanism}
\begin{figure}
    \centering
    \includegraphics[width=0.92\linewidth]{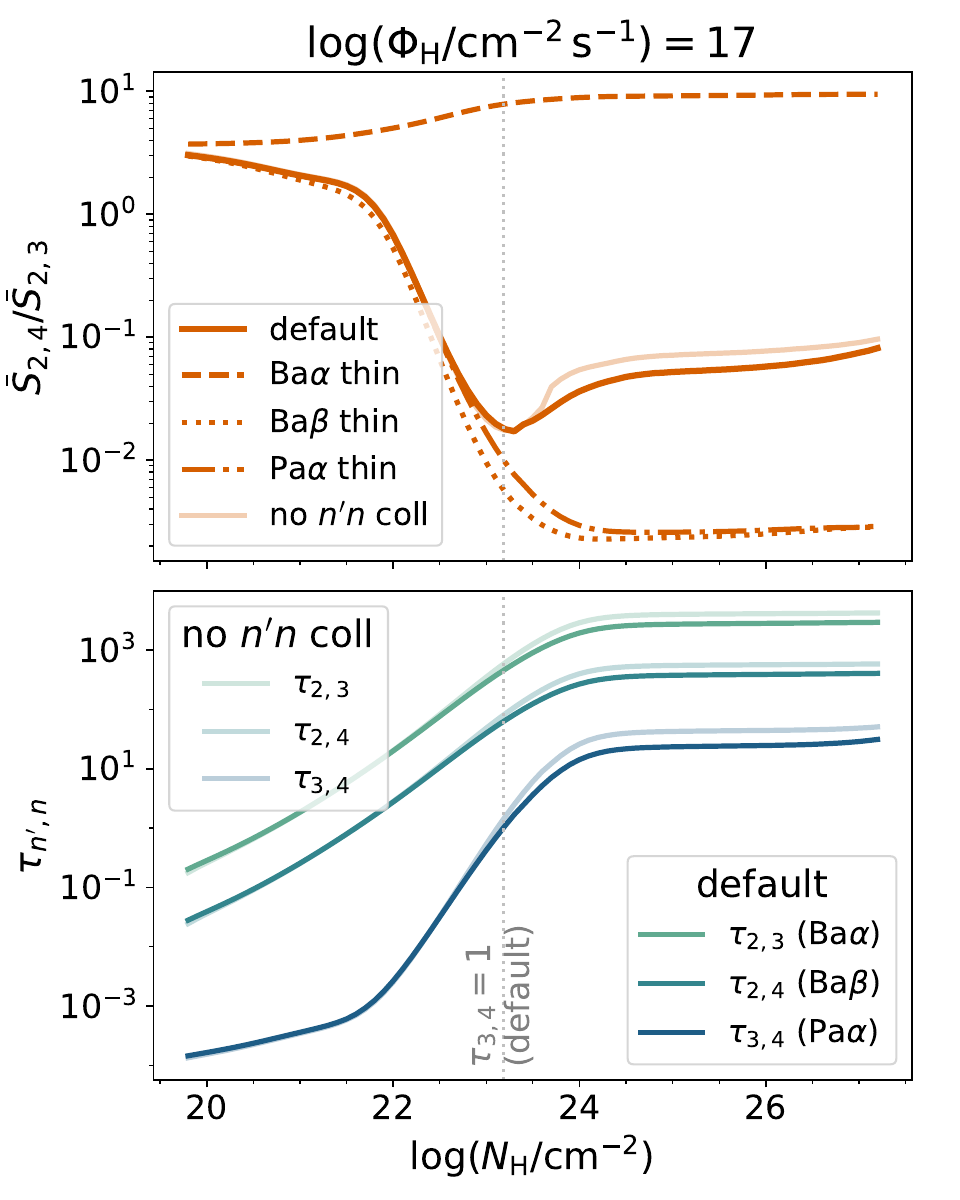}
    \caption{Top: default and artificially modified \textsc{cloudy} models at $\log(\Phi_{\rm H}/{\rm cm^{-2}\,s^{-1}})=17$. ``Ba$\alpha$ thin'', ``Ba$\beta$ thin'', and ``Pa$\alpha$ thin'' denote models with the escape probability of Balmer-$\alpha$, Balmer-$\beta$, and Paschen-$\alpha$ forced to unity, respectively; ``no $n'n$ coll'' denotes models without collisional (de-)excitation. Bottom: line-center optical depths of Ba$\alpha$, Ba$\beta$, and Pa$\alpha$ in the default models. The grey dotted line marks where Pa$\alpha$ becomes optically thick in the default models. Maintaining low $\bar{S}_{2,4}/\bar{S}_{2,3}$ requires Ba$\alpha$ trapping in the absence of Ba$\beta$ or Pa$\alpha$ trapping. }
    \label{fig:cloudy_test}
\end{figure}
We now fix $\log(\Phi_{\rm H}/\rm cm^{-2}\,s^{-1})=17$ to study the mechanism of the weak H$\alpha$ absorption, i.e., the cause of the low $\bar{S}_{2,4}/\bar{S}_{2,3}$ ratio in our photoionization models. We calculate four additional sets of models. In the first set, we artificially set the escape probability of all $3\to2$ transitions to unity, thereby disabling H$\alpha$ (also referred to as Balmer-$\alpha$ below, or Ba$\alpha$) line trapping. The second and third sets are similar, except that we keep the default H$\alpha$ escape probability but disable H$\beta$ ($4\to2$) and Paschen-$\alpha$ ($4\to3$; Pa$\alpha$ hereafter) line trapping, respectively. In the fourth set, we keep radiative physics unmodified but turn off collisional (de-)excitation, i.e., collisional transitions across different $n$ levels. The results are shown in Figure~\ref{fig:cloudy_test}. Compared to the default models with full physics, models without Ba$\alpha$ trapping stop showing any significant decrease in $\bar{S}_{2,4}/\bar{S}_{2,3}$, whereas models without either Ba$\beta$ or Pa$\alpha$ trapping develop even lower $\bar{S}_{2,4}/\bar{S}_{2,3}$ at high $N_{\rm H}$ (i.e., high $N_2$). Notably, the default and the Pa$\alpha$-thin models start to diverge where the line-center optical depth of Pa$\alpha$ reaches unity, which coincides with the minimum in $\bar{S}_{2,4}/\bar{S}_{2,3}$ in the default model. Models without collisional (de-)excitation are qualitatively similar to the default set.

These experiments suggest that a low $\bar{S}_{2,4}/\bar{S}_{2,3}$ is caused by different radiative trapping efficiencies of the $n=3$ versus $n=4$ levels. In the absence of any line trapping (except for Lyman lines, which are always optically thick in our case of interest), the $n\geq3$ level population would be dictated by the recombination cascade (and collisions to a lesser extent). This is the classical Case B scenario \citep{Hummer1987}; we have also performed single-zone \textsc{cloudy} calculations for this idealized case (more precisely, with the input command ``Case B Hummer no photoionization no Pdest''; not visualized here) for the same $\Phi_{\rm H}$ and found $\bar{S}_{2,4}/\bar{S}_{2,3}=3.6$, broadly similar to the behavior of the ``Ba$\alpha$ thin'' curve in Figure~\ref{fig:cloudy_test}. In contrast, in our default setup, Ba$\alpha$ has already become modestly optically thick at the lowest $N_{\rm H}$ and monotonically increases with the latter; see the bottom panel in Figure~\ref{fig:cloudy_test}. This enables radiative trapping to populate the $n=3$ state and thus to boost the source function $\bar{S}_{2,3}$. The boosting is efficient because there is no other permitted route of radiative decay for $n=3$. However, this is not the case for $n=4$. Although the optical depth of Ba$\beta$ is only a factor of a few lower than that of Ba$\alpha$, so both lines are optically thick in most of the parameter space here, $n=4$ has an alternative route of radiative decay, namely, Pa$\alpha$. As long as Pa$\alpha$ remains optically thin, the $n=4$ population will not build up efficiently. This explains why $\bar{S}_{2,4}/\bar{S}_{2,3}$ continues to decrease only until Pa$\alpha$ becomes optically thick; by then, all decay channels for $n=4$ are throttled, and $\bar{S}_{2,4}$ starts to grow and inflate the ratio $\bar{S}_{2,4}/\bar{S}_{2,3}$. The mechanism also explains why both the ``Ba$\beta$ thin'' and ``Pa$\alpha$ thin'' curves stay at low $\bar{S}_{2,4}/\bar{S}_{2,3}$ under high $N_{\rm H}$: the $n=4$ population is drained when either decay channel is artificially kept unrestricted. 

Much of the mechanism here has already been invoked to explain the large Balmer decrement of the broad Balmer emission lines in LRDs \citep{Yan2026}. Although here we focus on the source function rather than on the emission line flux, the relevant microphysical processes, i.e., radiative trapping and pumping, are identical. Fundamentally, when $n_{n'\ell'}\gg n_{n\ell}$ and when the $n'\ell'$ sublevels are $\ell$-mixed (both are valid in our parameter space), the denominator in Equation~(\ref{eq:source_function}) will be proportional to the total lower-level density $n_{n'}\equiv\sum_{\ell'}n_{n'\ell'}$, which is shared by $S_{2,4}$ and $S_{2,3}$. This implies that the source function ratio $S_{2,4}/S_{2,3}$ is proportional to the emissivity ratio between $4\to2$ and $3\to2$, which is directly related to the emission line flux ratio. Therefore, the shallow absorption of H$\alpha$ relative to H$\beta$ could be viewed as a strong Balmer decrement manifesting at the absorption trough. We note that the absorption ``decrement'' has to be more extreme than that of the broad emission luminosity to produce the inverted absorption EW ratio (Equation~\ref{eq:source_function_ratio}), so a strong broad emission decrement does not necessarily imply an inverted absorption ratio. 

We incidentally discuss the role of collisional (de-)excitation. In Figure~\ref{fig:cloudy_test}, when $\log(N_{\rm H}/\rm cm^{-2})\leq23$, the models with the process turned off (``no $n'n$ coll'') are almost identical to default. At higher $N_{\rm H}$, these modified models show slightly higher optical depth in the Balmer and Paschen lines than default. The comparison suggests that collisional effects serve as a net sink to the excited levels at high $N_{\rm H}$ and hardly affects the level population at low $N_{\rm H}$. At any rate, the ``no $n'n$ coll'' models also develop a low $\bar{S}_{2,4}/\bar{S}_{2,3}$, which indicates that the mechanism of weak H$\alpha$ absorption itself does not require collisional (de-)excitation. However, collisions could still positively populate the $n=2$ population in other density and radiation regimes (\citealt{Yan2026} at $n_{\rm H}\leq10^8{\rm~cm^{-3}}$; verified in controlled tests similar to this work, Z. Yan, private communication). 

\section{Curve of growth with source function}
\label{sec:curve_of_growth_source}
\begin{figure*}
    \centering
    \includegraphics[width=0.88\linewidth]{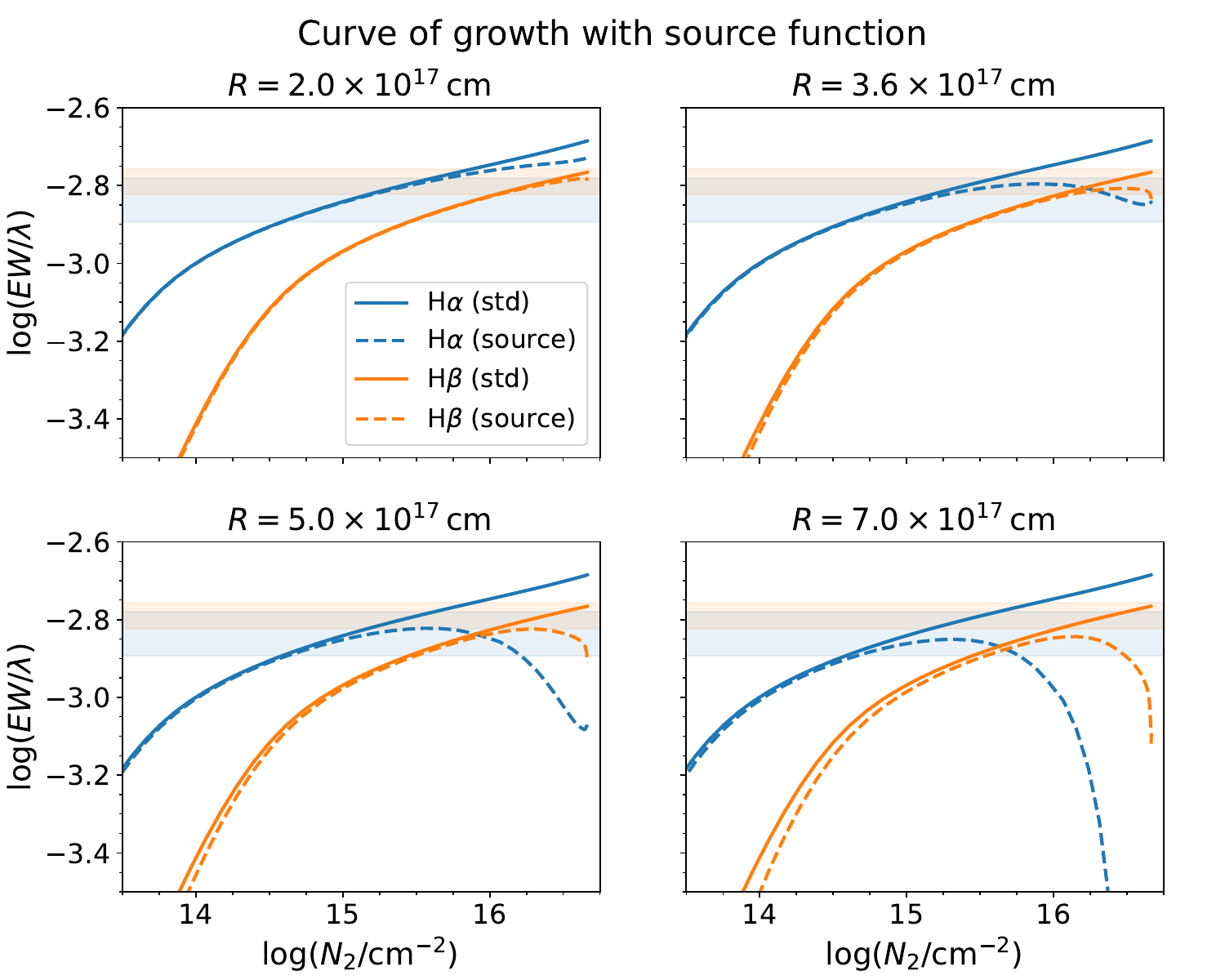}
    \caption{Curve of growth with source function. The solid curves are the standard curves of growth, indentical to the middle panel in Figure~\ref{fig:curve_of_growth}. The dashed curves are the source-corrected curves using the source function from \textsc{cloudy} models (with fixed $\log(\Phi_{\rm H}/{\rm cm^{-2}\,s^{-1}})=17$ and $b=100{\rm~km\,s^{-1}}$) and the observed broad-line luminosities of RUBIES-49140 (Equations~\ref{eq:EW}, \ref{eq:radius_inference}); c.f. Figure~\ref{fig:curve_of_growth}, where the dashed curves are only conceptual illustrations without microphysical modeling. Shaded horizontal bands are the observationally estimated EWs for H$\alpha$ (blue) and H$\beta$ (orange) for RUBIES-49140 with 1$\sigma$ uncertainties. Depending on the radius of the absorbing medium, the EW could strongly deviate from the standard at high column densities and could be weaker in H$\alpha$ than in H$\beta$. }
    \label{fig:curve_of_growth_cloudy}
\end{figure*}

In Section~\ref{sec:problem}, we motivated the modified curve of growth by including the source function in Equation~(\ref{eq:EW}), but we only naively shifted the curves in Figure~\ref{fig:curve_of_growth} for conceptual demonstration. Now, the photoionization models inform us of $\bar{S}_{2,3}$ and $\bar{S}_{2,4}$, which enables us to construct the modified curve of growth quantitatively. We choose the family of models at $\log(\Phi_{\rm H}/{\rm cm^{-2}\,s^{-1}})=17$ as an example and measure $\bar{S}_{2,3}$ and $\bar{S}_{2,4}$ for each model, i.e., each $\log(N_{\rm H}/{\rm cm^{-2}\,s^{-1}})$. Then, we use Equation~(\ref{eq:radius_inference}) to obtain $\langle S/I_{\rm in}\rangle_{2,3}$ and $\langle S/I_{\rm in}\rangle_{2,4}$. This requires knowing $L_{\rm in,2,3}$, $L_{\rm in,2,4}$, and $R$: we take observed values for the luminosities (see text below Equation~\ref{eq:radius}) and treat $R$ as a free parameter. Finally, we substitute $\langle S/I_{\rm in}\rangle_{2,3}$ and $\langle S/I_{\rm in}\rangle_{2,4}$ into Equation~(\ref{eq:EW}) to obtain the source-corrected EW, where we assume $C=1$ for simplicity.

Figure~\ref{fig:curve_of_growth_cloudy} shows the results. The standard curves of growth (solid) are identical to those in the middle panel in Figure~\ref{fig:curve_of_growth}. The curves that consider the source function (dashed) depart from the standard ones in non-trivial ways. For fixed values of $\bar{S}_{2,3}$ and $\bar{S}_{2,4}$, a small radius of the absorbing medium implies a weak correction term to the EW by Equation~(\ref{eq:radius_inference}), as shown in the top-left panel. However, if we assume larger radii, the source correction will strongly bend the curves. Importantly, the trough in $\bar{S}_{2,4}/\bar{S}_{2,3}$ in Figure~\ref{fig:cloudy_grid} implies that the H$\alpha$ EW could decrease more significantly than H$\beta$ at $\log(N_2/{\rm cm^{-2}})\sim15$--17. Indeed, the source-corrected H$\alpha$ curve starts to deviate from the standard one at $\log(N_2/{\rm cm^{-2}})\sim15$ across the four panels, and a crossing between the H$\alpha$ and H$\beta$ curves happens in three of the four panels here. After the crossing, we expect the same column density to produce $(EW/\lambda)_{\rm H\alpha}<(EW/\lambda)_{\rm H\beta}$, as observed in RUBIES-49140. When the radius of the medium is large, the EW could even become zero or negative for H$\alpha$ while staying large for H$\beta$; by then, only H$\beta$ would show an absorber at all. 

Regardless of radius, the deviations are strongest at the largest $N_2$, where radiative trapping effects (Section~\ref{subsec:mechanism}) are maximized. In other words, the source function impacts the curve of growth for gas media with high column densities. Therefore, compared to typical AGN broad-line regions, which have relatively low column densities, the source function is more plausibly non-negligible in the absorption analysis of LRDs. 

In Figure~\ref{fig:curve_of_growth_cloudy}, a radius of $R=3.6\times10^{17}{\rm~cm}$ for the absorbing medium could simultaneously explain the H$\alpha$ and H$\beta$ EWs at $\log(N_2/{\rm cm^{-2}})\sim16.3$--16.7. An independent estimate of the radius can be made by comparing the flux of the input radiation in the photoionization models and the UV-optical luminosity of the LRD, $L_{\rm UVopt}$ \citep[e.g.,][]{Juodzbalis2024}. The latter is uncertain for RUBIES-49140 due to the lack of observation at the rest-frame red optical, so we take $L_{\rm UVopt}\sim10^{44.5}{\rm~erg~s^{-1}}$ as a crude estimate, which corresponds to $\lambda L_\lambda$ at rest-frame $6000\rm~\AA$. Combined with the parameters of the models presented here, $T_{\rm eff}=20000{\rm~K}$ and $\log(\Phi_{\rm H}/{\rm cm^{-2}\,s^{-1}})=17$, we infer $R_L=6.5\times10^{17}{\rm~cm}$, broadly consistent with the radius inferred above. The values are larger than the photosphere radius inferred from the rest-optical continuum emission, $R_{\rm ph}\sim10^{16-17}{\rm~cm}$ \citep[e.g.,][]{deGraaff2025b,Sun2026}, which suggests a radially extended gas distribution outside the continuum photosphere. However, we caution that both $R$ and $R_L$ estimates are subject to large uncertainties from the photoionization models, including but not limited to the incident radiation and the gas spatial distribution. The quoted values of $N_2$ are similarly uncertain. As we have stated, the photoionization calculations serve as a proof of concept rather than quantitative predictions. 

\section{Summary and discussion}
\label{sec:summary}

Observations of the Balmer absorbers of LRDs have identified a problem: the absorption EW ratio of H$\alpha$ to H$\beta$ can be much lower than expected from the standard attenuation model. However, the standard attenuation model assumes that the absorbing medium does not emit light in its own right. Considering the source function of this medium can naturally resolve the tension. Based on this central idea, we propose a modified attenuation model that considers the source function (Equation~\ref{eq:attenuation_model}). Our main conclusions are as follows.

\begin{enumerate}
    \item Observations of some LRDs such as RUBIES-EGS-49140 show that $(EW/\lambda)_{\rm H\alpha}<(EW/\lambda)_{\rm H\beta}$. This poses a tension with the standard curve of growth, which predicts $(EW/\lambda)_{\rm H\alpha}>(EW/\lambda)_{\rm H\beta}$ for any thickness of the absorbing medium (Figure~\ref{fig:curve_of_growth}).
    \item Including the source function into the standard attenuation model scales down the predicted equivalent width (Equations~\ref{eq:attenuation_model}, \ref{eq:EW}). The correction can be different for H$\alpha$ and H$\beta$. 
    \item Proof-of-concept photoionization models using \textsc{cloudy} show that the source function of H$\alpha$ could be systematically stronger than that of H$\beta$ (Figure~\ref{fig:cloudy_grid}) due to radiative trapping, which operates efficiently when the absorbing medium is optically thick to H$\alpha$ but not very optically thick to Paschen-$\alpha$ (Figure~\ref{fig:cloudy_test}). 
    \item The source-corrected curve of growth could show inverted H$\alpha$-to-H$\beta$ EW ratios, consistent with observations (Figure~\ref{fig:curve_of_growth_cloudy}). The EWs are sensitive to the location of the absorbing medium. 
\end{enumerate}

Our results motivate a more general attenuation model than that commonly assumed in the literature (see Equation~\ref{eq:attenuation_model}). With the source function, the optical depth and covering factor measured from the standard attenuation model should not be taken at face value. Our conclusion aligns with the emerging picture that many LRD features share similarities with stellar phenomena and could be understood with physics in that context \citep[e.g.,][]{Naidu2025,Kido2025,Liu2025,Begelman2025,Nandal2026,Naidu2026}. Indeed, a stellar atmosphere has practically infinite optical depth for the Balmer lines, yet their EWs are fully governed by the source function rather than follow the standard curve of growth (which would predict infinity). 

All else being equal, the source function becomes increasingly important at high $N_2$ (Figure~\ref{fig:curve_of_growth_cloudy}). This suggests that the individual absorption EW need not monotonically increase with gas column density, complicating interpretations of observational correlations (or the lack thereof) involving the absorber EW. However, we may expect a correlation between the occurrence of an anomalous EW ratio and very high gas column densities. The latter could be independently constrained by other absorption lines or the Balmer break, although they do not necessarily trace the same gas column; we discuss this further in Section~\ref{subsec:balmer_break}. Clarifying the demographics of EW ratios will benefit from a broader sample of high-resolution, high signal-to-noise-ratio spectra.  

Our framework also suggests a solution to the multiple absorber problem in some LRDs, where H$\alpha$ may show absorption at an offset unmatched by any obvious absorption in H$\beta$, and vice versa \citep[e.g.,][]{Lin2026Egg}. We could conceptually invoke distinct absorbing media at different line-of-sight velocities. An H$\alpha$-only absorber could be explained by the standard curve of growth at a small optical depth; an H$\beta$-only absorber could be the strong-source limit of the shallow H$\alpha$ absorber, where a high optical depth could completely annihilate H$\alpha$ absorption (bottom-right panel of Figure~\ref{fig:curve_of_growth_cloudy}). Under this interpretation, the H$\alpha$ absorption no longer follows the gas at high column density that is traced by H$\beta$. This refines the existing idea that high-order Balmer absorbers may be sensitive to media with high column densities \citep[e.g.,][]{Yanagisawa2026b}. 

We focus on the EW in this paper and defer detailed models to the absorber profile to future works. In Equation~(\ref{eq:attenuation_model}), the source function likely varies with wavelength due to stratification in the absorbing medium. This variation will make the absorber deviate from a ``saturated'', boxy profile that follows from the standard attenuation model and poorly fits LRD observation \citep{Chen2026}; indeed, typical stellar absorption lines do not appear boxy. Velocity gradients in the stratified absorbing medium could produce the observed transition-dependent offsets \citep{Wang2026b}, as in models of stellar P Cygni lines \citep[e.g.,][for P Cygni itself]{vanBlerkom1978}. An important future application is to decode the source function at different depths by modeling the absorber profile, i.e., using the absorber as a tomogram of the gas layer. Similar methods have long been used to diagnose the stratification of stellar atmospheres or envelopes \citep[e.g.,][]{Fuhrmann1993,vanBlerkom1978}.

\subsection{Comparison to the partial covering factor scenario}

The absorption strength problem is also present in broad absorption line quasars \citep[BAL quasars; e.g.,][for Balmer lines]{Hall2002,Aoki2006,Leighly2025}. In that context, the unexpected ratios in absorption optical depths combined with saturated yet nonzero fluxes at the absorption line cores largely motivated the now-standard attenuation model, with a partial covering factor (e.g., \citealt{Barlow1997,Hamann2001}; see also, e.g., \citealt{deKool2002} for an inhomogeneous covering factor scenario). Our model is similarly motivated by anomalous absorption strength ratios and is compatible with the standard attenuation model (see Equation~\ref{eq:attenuation_model}). However, our framework is different in two aspects. First, whereas the standard model attributes the phenomenon purely to the gas geometrical distribution, we ascribe it at least partly to self-emission from the absorbing gas. Second, since H$\alpha$ and H$\beta$ are absorbed by the same $n=2$ level, it is plausible that they should share the same covering factor regardless of the geometry of the absorbing medium (but see below); if so, it is not clear to us how partial covering alone can explain the inverted $EW/\lambda$ ratio seen in RUBIES-49140 (Section~\ref{sec:problem}). On the other hand, the source functions can naturally differ between H$\alpha$ and H$\beta$ due to their different upper levels. 

A caveat to the above reasoning about the same covering factor for H$\alpha$ and H$\beta$ is if the total emission (before passing through the absorbing medium) arises from more than a single gas component. For example, the continuum-emitting gas and the broad-line-emitting gas may be separate and have different covering factors \citep[][in the context of BAL quasars]{Leighly2025}. The broad H$\alpha$ and H$\beta$ emission also possibly arise from different gas components, as discussed in, e.g.,  \citet{Wang2026b}. These scenarios could enable different covering factors for H$\alpha$ and H$\beta$ to explain the observed anomalous absorption despite their common lower level. However, the source function model proposed here is known to operate in stellar atmospheres and stellar winds and, if further confirmed in LRDs (or even potentially in other types of AGNs), would imply a qualitatively different picture of the absorbing gas than partial covering. Further theoretical studies and independent observational constraints are needed to distinguish these two models.

\subsection{The Balmer absorbers and the Balmer break}
\label{subsec:balmer_break}
The Balmer absorber has been suggested to relate to the Balmer break of LRDs both theoretically \citep{Inayoshi2025,Ji2025} and observationally \citep[e.g.,][]{Matthee2026,Wang2026b}: the same medium that produces the Balmer line absorption may also be responsible for the observed Balmer break in a subset of LRDs. The bound-free absorption cross-section at the Balmer limit is $\sigma_{2\infty}=1.38\times10^{-17}{\rm~cm^{-2}}$ \citep{Karzas1961}. For a given absorbing medium, the optical depth at the Balmer limit $\tau_{2\infty}=\sigma_{2\infty}N_2$ is less than that at the center of H$\alpha$ or H$\beta$; cf. Equation~(\ref{eq:optical_depth_center}). Therefore, while the formation of both Balmer line absorption and the Balmer break require high $N_2$, a high optical depth to Balmer lines is a necessary but not sufficient condition for a given absorbing medium to imprint a Balmer break on an otherwise featureless continuum \citep[see also][]{Chen2026}. In particular, the column densities suggested by our photoionization models, $\log(N_2/{\rm cm^{-2}})\sim15$--17 (but with uncertainties that we have not quantified) are at best marginally optically thick at the Balmer limit. This suggests two possibilities: either the absorbing medium has higher $N_2$ than inferred from our simplistic models, or that the Balmer break forms at a gas layer deeper in than the line-absorbing medium, possibly within the broad-line region itself or toward the optical continuum photosphere \citep{Liu2025}. High-resolution spectroscopy combined with kinematic modeling for the Balmer break spectral region and the Balmer absorbers could distinguish the two scenarios: a distinct origin for the break may be favored if the break smoothness requires significantly larger turbulence velocity than allowed by the line absorber \citep[see][where the break required a microturbulence at least as large as the absorber width]{Ji2025}. 

\subsection{A potential new method for mass estimation}

The radius dependence of the source-corrected curve of growth suggests an avenue to constrain the radius of the absorbing medium $R$, which, under additional dynamical assumptions, could lead to a mass estimate. If the absorbing medium is virialized, the dynamical mass within the medium will be given by
\begin{equation}
    M = {b^2R\over G} = 7.5\times10^{5}~M_\odot\left({b\over100{\rm~km\,s^{-1}}}\right)^2{R\over10^{18}{\rm~cm}}\,, \label{eq:mass}
\end{equation}
and the implied Eddington ratio is
\begin{gather}
    \lambda_{\rm Edd} \sim {L_{\rm UVopt}\over L_{\rm Edd}(M)} \nonumber \\
    = 9\times10^0{L_{\rm UVopt}\over10^{45}{\rm~erg\,s^{-1}}}\left({b\over100{\rm~km\,s^{-1}}}\right)^{-2}\left({R\over10^{18}{\rm~cm}}\right)^{-1}\,.
\end{gather}
We caution that, in Section~\ref{sec:curve_of_growth_source}, we have not quantified the uncertainty in $R$. Within the framework of \textsc{cloudy}, the radius likely depends on the incoming radiation (for which most literature works, including us, assume a smooth continuum, whereas potential lines could affect the results by pumping; see Appendix~\ref{app:test_Lya}) and the extent and density profile of the gas medium. Broader physical effects beyond the scope of \textsc{cloudy}, e.g., a general velocity field \citep[such as the outflows indicated by the commonly blueshifted absorbers, e.g.,][]{Matthee2026,Juodzbalis2026,Yanagisawa2026b,Chen2026}, scattering either in physical or frequency space, inhomogeneity, and a non-spherical gas distribution, may affect both the predicted source function and the inferred location of the absorbing medium. On the dynamical side, we have not justified the virial assumption in Equation~(\ref{eq:mass}). However, both radiative transfer and dynamics are amenable to future, more self-consistent modeling. By Equation~(\ref{eq:mass}), we motivate the source-function-mediated absorber as a potential mass probe that awaits a more detailed joint analysis between theory and data. Overall, we believe that the source function provides an interpretation of the suppressed H$\alpha$/H$\beta$ absorption ratio in LRDs, but more work is required to understand the range of physical conditions under which this mechanism operates (including the potential for multiple physical solutions). Answering this would turn the Balmer absorbers from a puzzle into a clue about the physical conditions in LRDs.

\begin{acknowledgments}
HL thanks Yilun Ma for discussions on absorption lines in stellar atmospheres, Zu Yan for tests of collisional (de-)excitation in the setup of \citet{Yan2026}, and Kohei Inayoshi and Ignas Juod{\v{z}}balis for constructive discussions. Support for US investigators in Program \# JWST-AR-10935.002-A was provided by NASA through a grant from the Space Telescope Science Institute, which is operated by the Association of Universities for Research in Astronomy, Inc., under NASA contract NAS5-03127.
B.W. acknowledges support provided by NASA through Hubble Fellowship grant HST-HF2-51592.001 awarded by the Space Telescope Science Institute, which is operated by the Association of Universities for Research in Astronomy, In., for NASA, under the contract NAS 5-26555.
This work is based in part on observations made with the NASA/ESA/CSA James Webb Space Telescope. The data were obtained from the Mikulski Archive for Space Telescopes at STScI. These observations are associated with program \# 8047.

\end{acknowledgments}

\software{Astropy \citep{astropy2013,astropy2018,astropy2022}, Claude Code \citep{Anthropic2025}, cloudy \citep{Gunasekera2025}, Matplotlib \citep{Hunter2007}, Numpy \citep{Harris2020}, OverCite \citep{Shariat2026}}

\appendix

\section{Effect of Stark and self-broadening on the curve of growth}
\label{app:stark}
\begin{figure}
    \centering
    \includegraphics[width=0.99\linewidth]{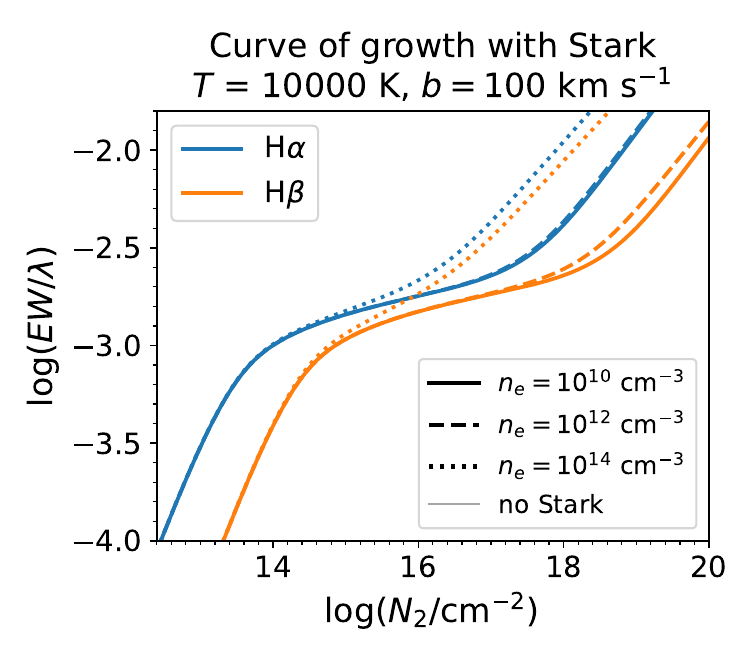}
    \caption{Curve of growth with the Stark broadening effect but without the source function. The curves at $n_e=10^{10}{\rm~cm^{-3}}$ are visually indistinguishable from the curves without the Stark effect. Stark broadening tends to shrink the difference between H$\alpha$ and H$\beta$, but it only significantly operates at $n_e>10^{12}{\rm~cm^{-3}}$ and still gives $(EW/\lambda)_{\rm H\alpha}>(EW/\lambda)_{\rm H\beta}$.   }
    \label{fig:curve_of_growth_stark}
\end{figure}

The curve of growth in the main text considers only Doppler broadening and radiative (natural) damping. On the other hand, in stellar atmospheres, the Stark effect and self-broadening are important collisional broadening mechanisms for the Balmer lines \citep[e.g.,][]{Ali1965,Griem1974,Fuhrmann1993,Barklem2000}. In a dense medium, collisions between the line-emitting hydrogen atom with charged particles (Stark) or with another neutral hydrogen atom (self) contribute to the spectral wing. This could dominate over the natural damping wing and change the standard curve of growth even without considering the source function. However, both effects are negligible in the typical gas density considered here ($n_{\rm H}\sim10^{10}{\rm~cm^{-3}}$) and do not by themselves resolve the weak H$\alpha$ absorption problem.

Figure~\ref{fig:curve_of_growth_stark} shows the curve of growth considering the Stark effect but without the source function, assuming a uniform gas temperature of 10000~K (using 5000~K or 20000~K instead would give a $<3\%$ difference). For H$\alpha$ and H$\beta$ respectively, we use the tabulated Stark-broadened profiles in \citet{Stehle1999} and convolve them with the same Voigt profiles that give the standard curves of growth in Figure~\ref{fig:curve_of_growth}. Overall, the Stark effect broadens the far wings and thus enhances the EW at high $N_2$. Its effect strongly depends on the electron density: it has negligible impact on the curve of growth at $n_e=10^{10}{\rm~cm^{-3}}$ but can boost the EW by a factor of several in the figure at $n_e=10^{14}{\rm~cm^{-3}}$. The effect tends to broaden H$\beta$ more strongly than H$\alpha$, in the direction of alleviating the weak H$\alpha$ absorption problem of LRDs. However, our photoionization models in Section~\ref{sec:microphysics} are in the regime of relatively low-density, predominantly neutral media ($n_e\ll n_{\rm H}=10^{10}{\rm~cm^{-3}}$), where this effect is suppressed. Even if the absorbing medium is in fact dense enough for the Stark effect to dominate, we still have $(EW/\lambda)_{\rm H\alpha}/(EW/\lambda)_{\rm H\beta}\geq1.2$; this is an analytical limit independent of temperature or density \citep{Stehle1999}. Therefore, the Stark effect alone cannot produce the inverted EW ratio observed in LRDs.

Self-broadening produces a Lorentzian profile and is directly comparable to radiative damping (unlike Stark broadening). The half width at half maximum of self-broadening is proportional to the neutral hydrogen density $n_{\rm HI}$ and is given by \citep[e.g.,][]{Barklem1997}
\begin{equation}
    w_{\rm self} = n_{\rm HI}\left({4\over\pi}\right)^{\alpha/2}\Gamma\left({4-\alpha\over2}\right)\bar{v}\sigma(v)\,,
\end{equation}
where $\Gamma(\cdot)$ is the Gamma function, $\bar{v}$ is the mean collision speed between two hydrogen atoms, $\sigma(v)\propto v^{-\alpha}$ is the broadening cross-section. Using Table~3 from \citet{Barklem2000} and taking a characteristic value of $\bar{v}=10{\rm~km\,s^{-1}}$, we obtain $w_{\rm self,H\alpha}/n_{\rm HI} = 3.2\times10^{-8}{\rm~s^{-1}\,cm^3}$ and $w_{\rm self,H\beta}/n_{\rm HI} = 6.3\times10^{-8}{\rm~s^{-1}\,cm^3}$. For H$\alpha$ and H$\beta$, $0<\alpha<1$, so the result is not highly sensitive to $\bar{v}$. In comparison, the radiative damping of H$\alpha$ and H$\beta$ is given by $w_{\rm rad,H\alpha}=2.8\times10^8{\rm~s^{-1}}$ and $w_{\rm rad,H\beta}=2.5\times10^8{\rm~s^{-1}}$ \citep[using Einstein A values from][]{Wiese2009}. Therefore, self-broadening would become significant only when $n_{\rm HI}>10^{15}{\rm~cm^{-3}}$. Even if self-broadening does dominate over natural damping, the curve of growth will still show $(EW/\lambda)_{\rm H\alpha}>(EW/\lambda)_{\rm H\beta}$ despite the fact that $w_{\rm self,H\alpha}<w_{\rm self,H\beta}$: in the large-optical-depth limit, $(EW/\lambda)_{n',n}\propto \lambda_{n',n}(w_{n',n}f_{n',n}N_{n'})^{1/2}$, and the much higher oscillator strength of H$\alpha$ more than compensates for its somewhat smaller $w_{\rm self}$.  

\section{Tests for modification to \textsc{cloudy}}
\label{app:cloudy_modification}
\begin{figure}
    \centering
    \includegraphics[width=0.99\linewidth]{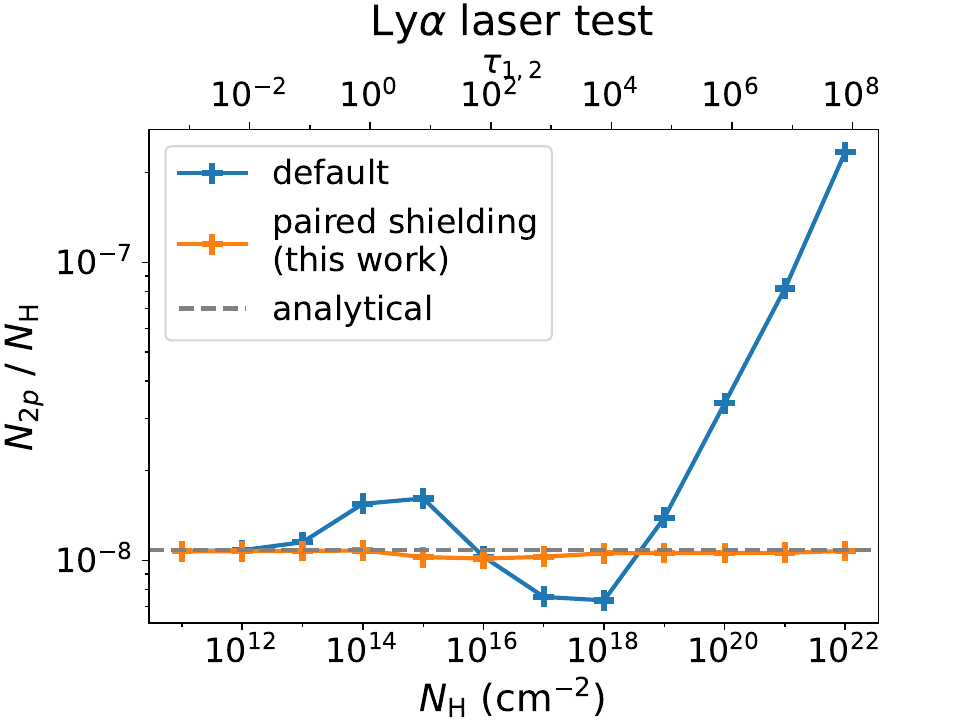}
    \caption{Ly$\alpha$ laser test for the default shielding function and our modified version (paired shielding). The top axis shows the optical depth at the line center of Ly$\alpha$. The default treatment progressively overpredicts the column density $N_{2p}$ as the optical depth increases. }
    \label{fig:shield_test_lyalaser}
\end{figure}
\begin{figure}
    \centering
    \includegraphics[width=0.99\linewidth]{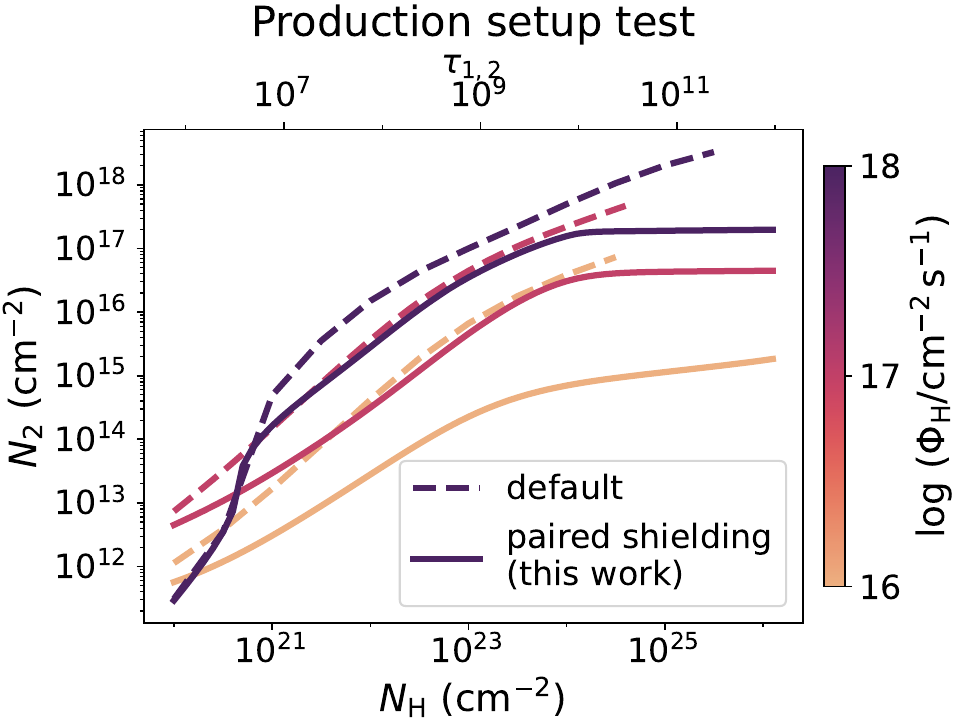}
    \caption{Test of the default \textsc{cloudy} shielding function and our modification using the same setup as in Section~\ref{subsec:cloudy}. The default treatment overpredicts the column density $N_2$ even when the incident spectrum is smooth. }
    \label{fig:shield_test_blackbody}
\end{figure}
As we discussed in Section~\ref{subsec:cloudy}, the default treatment in \textsc{cloudy} for the continuum shielding of each line is not paired with the escape probability formula of that very line. We now set up test cases to show the effects on its calculation. All tests in this section are reproducible from a Zenodo repository \citep{Liu2026software}.

We design our first test to isolate the radiative coupling of the $n=2$ level of hydrogen. We set a narrow continuum incident radiation confined to the Ly$\alpha$ wavelength, with a mean intensity given by $4\pi J_\nu=10^{-8}{\rm~erg\,s^{-1}\,cm^{-2}\,Hz^{-1}}$. Incoming radiation in all other wavelengths is zeroed. This Ly$\alpha$ laser illuminates a plane-parallel slab of gas in a closed geometry, but the line photons are allowed to escape from either side of the slab (the ``sphere'' command without ``static''). The gas has a uniform density $n_{\rm H}=10^{10}{\rm~cm^{-3}}$ and microturbulence $b=100{\rm~km\,s^{-1}}$. The elemental abundance is kept to default, but molecular chemistry is turned off. We hold the gas temperature constant at 2.8~K to suppress collisional processes. Under these conditions, the 2p population is controlled by the balance between radiative pumping and Ly$\alpha$ escape at each location, with (the medium being neutral throughout)
\begin{equation}
    n_{2p} = {g_{2p}\over g_{1s}}{2P_{\rm pen}(\tau)\over P_{\rm esc,in}(\tau)+P_{\rm esc,out}(\tau_s-\tau)}n_\gamma n_{\rm H}\,,
\end{equation}
where $g_{2p}/g_{1s}=3$, $n_\gamma=3.60\times10^{-9}$ is the occupation number of the Ly$\alpha$ laser in our setup, $P_{\rm pen}$ is the penetration probability (also known as the shielding factor), and $P_{\rm esc,in}$ and $P_{\rm esc,out}$ are the inward and outward escape probabilites. Here, $P_{\rm pen}$ and $P_{\rm esc,in}$ are functions of $\tau$, the optical depth towards the illuminated surface, whereas $P_{\rm esc,out}$ is a function of $\tau_s-\tau$, the optical depth towards the dark side; $\tau_s$ is the total optical depth of the slab. In particular, if the three probabilities have the same functional form, $P_{\rm pen}(\tau')=P_{\rm esc,in}(\tau')=P_{\rm esc,out}(\tau')\equiv P(\tau')$, then one obtains
\begin{equation}
    N_{2p} = {g_{2p}\over g_{1s}}n_\gamma N_{\rm H}\int_0^{\tau_s}{2P(\tau')\over P(\tau')+P(\tau_s-\tau')}{d\tau'\over\tau_s} = {g_{2p}\over g_{1s}}n_\gamma N_{\rm H}\,, \label{eq:laser_analytical}
\end{equation}
independent of the specific functional form of $P(\tau')$. 

Physically, the shared functional form expresses time-reversal invariance: the probability that a photon from the surface reaches a given depth, including all intermediate scatterings, equals the probability that a photon emitted from that depth reaches the surface. We term our treatment ``paired shielding'' since it pairs the shielding formula to the corresponding escape probability for each line. In contrast, in the default setup of \textsc{cloudy}, $P_{\rm esc,in}$ and $P_{\rm esc,out}$ have the same functional form for each line, which is nevertheless different from $P_{\rm pen}$. In particular, the default adopted formula \citep[adapted from][]{Federman1979} has $P_{\rm pen}(\tau')\propto \tau'^{-1/2}$ at large optical depths, whereas the default escape probabiliy formula for Ly$\alpha$ goes asymptotically $\propto\tau'^{-1}$ \citep[][]{Hummer1980}. This inconsistency could inflate the predicted $N_{2p}$ by a factor that scales with $\tau_s^{1/2}$. 

We run the above test setup for a range of $N_{\rm H}$. Figure~\ref{fig:shield_test_lyalaser} shows the test result. The paired shielding implementation in this work reproduces Equation~(\ref{eq:laser_analytical}) (grey dashed line) to within 10\%. In contrast, while the default treatment agrees with the analytical prediction in the optically thin limit, where all probabilities approaches unity, it severely overpredicts $N_{2p}$ at the high-optical depth end. There, the blue curve scales approximately as $N_{2p}/N_{\rm H}\propto N_{\rm H}^{1/2}$, as expected from our discussion above.

The laser incident radiation might appear artificial, but many existing applications of \textsc{cloudy} could enter the same regime even when assuming a continuum incoming spectrum. The Ly$\alpha$ intensity in our laser test is not very strong, with $4\pi J_\nu=10^{-8}{\rm~erg\,s^{-1}\,cm^{-2}\,Hz^{-1}}$. The same Ly$\alpha$ intensity is achieved by an undiluted blackbody at $T_{\rm eff}=6100{\rm~K}$, a diluted blackbody at $T_{\rm eff}=20000{\rm~K}$ with $\log(\Phi_{\rm H}/\rm cm^{-2}\,s^{-1})=16.9$ (i.e., an ionization parameter $\log U=-3.6$ assuming $n_{\rm H}=10^{10}{\rm~cm^{-3}}$), or a typical parameterized AGN spectrum at $T_{\rm BB}=10^5{\rm~K}, \alpha_{\rm OX}=-1.4, \alpha_{\rm UV}=-0.5, \alpha_{\rm X}=-1.0$ normalized by $\log(\Phi_{\rm H}/\rm cm^{-2}\,s^{-1})=17.6$ (i.e., $\log U=-2.9$ assuming $n_{\rm H}=10^{10}{\rm~cm^{-3}}$) (see documentation in \textsc{cloudy} for the definition). To test the impact on our production runs, we repeat the calculations in Section~\ref{subsec:cloudy} but using the default shielding formula instead, shown in Figure~\ref{fig:shield_test_blackbody}. Compared to models with our paired shielding treatment, which are the same as Section~\ref{subsec:cloudy}, the default treatment can often predict $N_2$ to be higher by one or two orders of magnitude. This would inflate the optical depth of the Balmer lines and continuum by the same amount at a given $N_{\rm H}$. 

The implications for downstream inference, such as the emergent continuum shape and line intensities (especially the Balmer break and Balmer transitions, which directly depend on $N_2$), are beyond the scope of this appendix. We suggest further tests in the community on existing \textsc{cloudy} models when the incident spectrum is non-negligible at the frequencies of Lyman lines -- including smooth continua -- and when the medium has a high H\RNum{1} column density for the difference between the shielding and escape treatment to manifest itself. 

\section{Test for the sensitivity to L\MakeLowercase{y}$\alpha$ pumping in production run}
\label{app:test_Lya}
\begin{figure}
    \centering
    \includegraphics[width=0.99\linewidth]{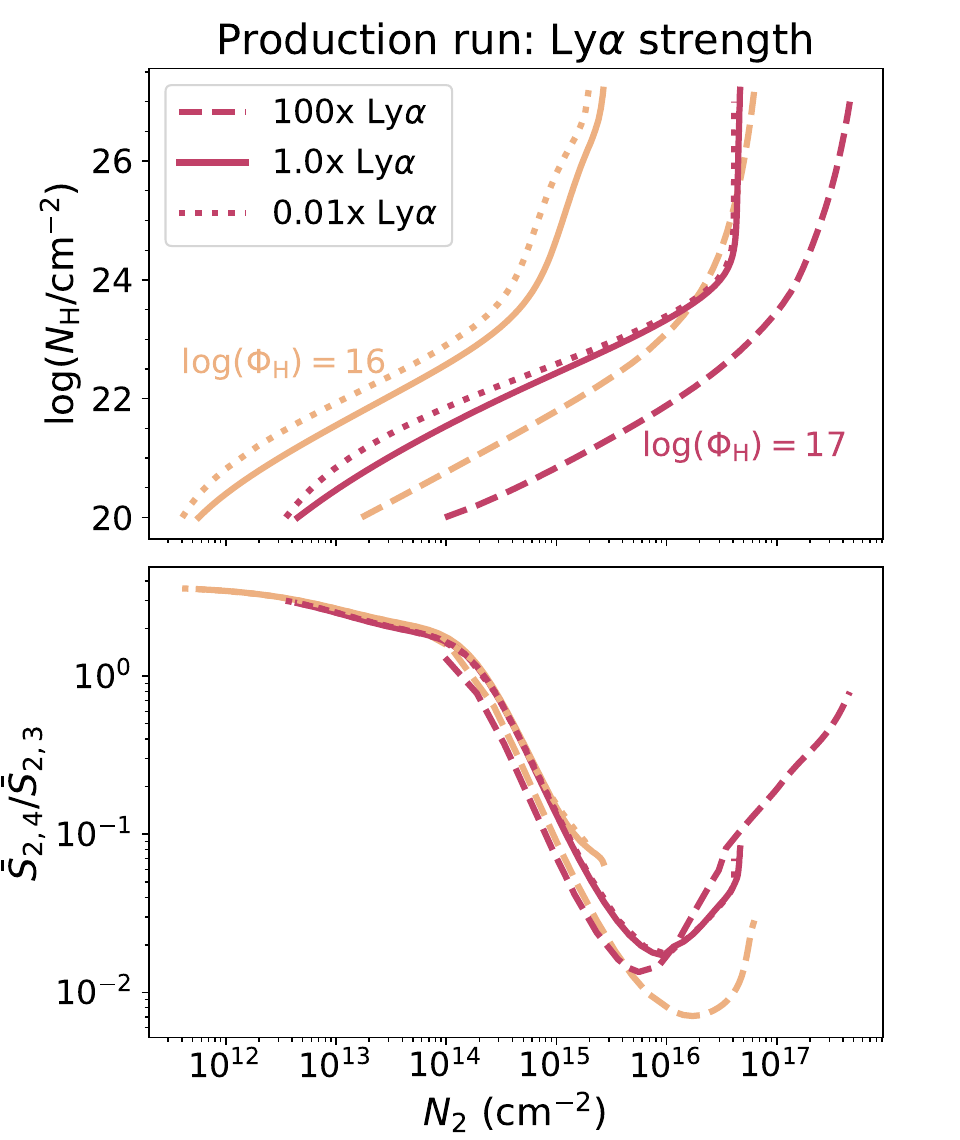}
    \caption{Production run with different Ly$\alpha$ strength in the incident spectrum. Light and dark curves correspond to $\log(\Phi_{\rm H}/{\rm cm^{-2}\,s^{-1}})=16$ and 17, respectively. The intensity at the Ly$\alpha$ wavelength is scaled by 100 (dashed) or 0.01 (dotted) compared to the fiducial, smooth continuum (solid). The same $\Phi_{\rm H}$ and $N_{\rm H}$ predicts different $N_2$ for the different assumed Ly$\alpha$ strength, indicating the effect of pumping by the incident Ly$\alpha$ line despite the same continuum spectrum. }
    \label{fig:test_Lya}
\end{figure}
As Appendix~\ref{app:cloudy_modification} shows, our paired shielding modification to \textsc{cloudy} predicts lower $N_2$ than the default treatment in the production run. With our modified \textsc{cloudy}, we now assess whether Ly$\alpha$ pumping by the incident continuum is still important and how sensitive the gas microphysical state depends on the strength of Ly$\alpha$ in the incoming spectrum. Different from Appendix~\ref{app:cloudy_modification}, which is concerned with computational consistency, this section investigates physical uncertainty. We deem the incoming Ly$\alpha$ intensity uncertain because the radiation entering the absorbing medium is likely already reprocessed by gas at smaller radii: the very gas that produces broad Balmer lines very likely also produces Lyman lines in emission or absorption. On this account, the assumption of a blackbody incident spectrum is very crude. 

Therefore, we perform additional \textsc{cloudy} runs with the identical setup as in Section~\ref{subsec:cloudy}, except that we simulate a Ly$\alpha$ emission or absorption line in the incoming spectrum by modifying the incoming radiation intensity exactly at the line wavelength. Other wavelengths remain unchanged. The results are shown in Figure~\ref{fig:test_Lya}. Compared to the fiducial models, scaling the Ly$\alpha$ incident intensity by a factor of 0.01 typically results in a factor-of-two reduction in $N_2$ at a given $N_{\rm H}$ and $\Phi_{\rm H}$, although the difference is less visible at $\log(\Phi_{\rm H}/{\rm cm^{-2}\,s^{-1}})=17$ at the high-$N_{\rm H}$ end. This demonstrates that, in our fiducial blackbody continuum with the paired shielding treatment, Ly$\alpha$ pumping by the incident radiation still significantly contributes to the creation of the $n=2$ hydrogen. On the other hand, assuming a Ly$\alpha$ incident radiation enhanced by a factor of 100 boosts $N_2$ by one to two orders of magnitude. The shape of the $\bar{S}_{2,4}/\bar{S}_{2,3}$ curve is also perturbed, albeit less dramatically: characteristically a factor of 2 difference for $N_2\gtrsim10^{15}{\rm~cm^{-2}}$ between the $100$x Ly$\alpha$ run and the fiducial run when their $N_2$ coverages overlap. Notably, the enhanced Ly$\alpha$ run at $\log(\Phi_{\rm H}/{\rm cm^{-2}\,s^{-1})}=16$ achieves $\bar{S}_{2,4}/\bar{S}_{2,3}<10^{-2}$ at $N_2\sim10^{16}{\rm~cm^{-2}}$, a column density beyond the reach of the fiducial models at the same ionizing flux at least for $\log(N_{\rm H}/{\rm cm^{-2}})\leq27$. Therefore, contrary to Figure~\ref{fig:cloudy_grid} might appear to suggest, producing low $\bar{S}_{2,4}/\bar{S}_{2,3}$ does not necessarily require a high $\Phi_{\rm H}$. 

We conclude that the uncertainties in the incident spectral lines, notably Ly$\alpha$, may affect the quantitative predictions of the models, e.g., the value of the ionizing photon flux that leads to shallow H$\alpha$ absorption. The qualitative trend remains robust as the radiative trapping machanism of H$\alpha$ persists (Section~\ref{subsec:mechanism}). 

\bibliography{bib}{}
\bibliographystyle{aasjournalv7.1}
\end{CJK*}

\end{document}